\documentstyle[12pt]{article}
\begin{document}
\title{The Holistic Universe and the Variation of the
Fine Structure Constant}
\author{B.G. Sidharth$^*$\\ Centre for Applicable Mathematics \& Computer Sciences\\
B.M. Birla Science Centre, Hyderabad 500 063 (India)}
\date{}
\maketitle
\footnotetext{E-mail:birlasc@hd1.vsnl.net.in}
\begin{abstract}
Inaba recently used a simple model to suggest that Quantum Theory can result from
a fluctuation in the cosmos. In this note we confirm this conclusion from a
different and more general point of view. We then argue that this provides an
explanation for the recently observed variation of the fine structure constant.
\end{abstract}
\section{Introduction}
Recently Inaba \cite{r1} has argued that a fluctuation in the Robertson-Walker
geometry yields a random motion for a particle in the universe, which latter
is equivalent to the usual Quantum Theory. He argues that this may be suggestive
of a manifestation of Mach's principle. Inaba's simple model is based on the work of
Santamoto \cite{r2,r3,r4,r5}, who tried to
give a geometrical interpretation for Quantum Theory. The object of the present
note is to confirm this result, from a different and more general viewpoint,
namely, in terms of fluctuations in the universe.
\section{Fluctuations}
We first observe that it is not surprising that Quantum Theory should be the
effect of fluctuations in the universe as a whole. In fact as pointed out
\cite{r6,r7,r8}, the fluctuation in the mass of a
typical elementary particle, for example the pion,
due to the fluctuation $\sim \sqrt{N}$ of the particle number $N \sim 10^{80}$
is given by
$$\Delta m \approx \frac{G\sqrt{N}m^2}{c^2R}$$
Whence
\begin{equation}
(\Delta mc^2)T = \frac{G\sqrt{N}m^2}{R}T = \frac{G\sqrt{N}m^2}{c}\label{e1}
\end{equation}
where $T$ is the age of the universe and $R$ its radius, which equals $cT$.\\
Not only is the right side of equation (\ref{e1}) the reduced Planck constant
$\hbar$, in the order of magnitude sense, but also equation (\ref{e1}) itself
is an expression of the Uncertainity relation
$$\Delta E \Delta t \approx \hbar .$$
Equation (\ref{e1}) again suggests the origin of Quantum Theory in
cosmic fluctuations.\\
Before proceeding it is worth mentioning that the above
line of reasoning leads to a cosmology which deduces from theory all the
supposedly miraculously coincidental large number relations of Dirac as also
the empirical inexplicable Weinberg formula which relates the mass of the
elementary particle, the pion to large scale parameters like the Hubble
Constant \cite{r9,r10},
$$m_\pi \approx \left(\frac{H\hbar^2}{Gc}\right)^{1/3}$$
The cosmological scheme is also
consistent with the subsequent discovery that not only is the universe
not decelerating but is actually accelerating and will expand for ever
\cite{r11,r12,r7,r8}.\\
Inaba deduces for a nearly flat Robertson-Walker universe from a minimum average curvature principle,
the Hamilton-Jacobi equation for a single particle,
\begin{equation}
\partial_t S + \frac{1}{2m} g^{\imath j}(\nabla_\imath S) + V -
\alpha R = 0.\label{e2}
\end{equation}
where the curvature $R$ is given by
$$R = R^{(b)} + R' ; \quad \quad R^{(b)} = 6 \left(\frac{\dot{a}}{a} + \frac{\ddot{a}^2}{a^2}\right),$$
$R'$ being the fluctuation effect, $R^{(b)}$ being the curvature in the standard
Robertson-Walker geometry.\\
Equation (\ref{e2}) leads by the standard Madelung-Bohm or Nelson Theory to
the Schrodinger equation
\begin{equation}
\imath \hbar \partial_t \psi = \frac{\hbar^2}{2m}\Delta \psi + V \psi -
\frac{\hbar^2}{4m} R^{(b)}\label{e3}
\end{equation}
Inaba then argues that (\ref{e3}) is indeed the Quantum Mechanical equation
in the classical Robertson-Walker geometry - it is the perturbation $R'$ in
the Robertson-Walker geometry that lead to (\ref{e3}).\\
We can justify the above conclusion as follows:
We first observe that in the random motion of $N$ particles, $l$ the fluctuation
in the length is given by
\begin{equation}
l \approx \frac{R}{\sqrt{N}}\label{e4}
\end{equation}
What is very interesting is that using for $R$ the actual radius of the universe
$\sim 10^{28}cm$, and for $N$ the actual number of particles in the universe,
(\ref{e4}) reduces to the well known Eddington formula, $l \sim 10^{-12}cm$ being the Compton
wavelength of a typical elementary particle, like the electron.\\
Further the diffusion equation describing the motion of a particle with
position given by $x(t)$ subject to random corrections is given by the well
known equation
$$| \Delta x | = \sqrt{< \Delta x^2 >} \approx \nu \sqrt{\Delta t},$$
where the diffusion constant $\nu$ is related to the mean free path $l$ and
the mean velocity $v$
\begin{equation}
\nu \approx lv\label{e5}
\end{equation}
Identifying $l$ of equation (\ref{e5}) with that in (\ref{e4}), we,  as
in the case of Nelson's derivation, arrive at the Hamilton-Jacobi
equation (\ref{e2}) and thence the Schrodinger equation (\ref{e3})
\cite{r13,r14,r15,r7}. Incidentally, this also provides a rationale to the
otherwise adhoc identification in Nelsonian theory viz.,
$$\nu = \hbar/m$$
Thus using the equations of Brownian Motion in the context of all the particles
in the universe, we arrive at the same equations
(\ref{e2}) and (\ref{e3}) of Inaba based on a minimum curvature principle and
Santamoto's geometric Quantum Mechanics.\\
Infact one can look upon the above results in terms of the fluctuation of the
metric itself. In Santamoto's original formulation \cite{r2,r3,r4,r5},
the geometry is Weyl's guage invariant geometry, where there is no invariant length
and infact we have
\begin{equation}
\delta l^2 \sim l^2 \delta g_{\imath k}\label{e6}
\end{equation}
It must be stressed that (\ref{e6}) is valid for arbitrary vectors $A^\imath$,
in which case $l$ would be their length.\\
Using the usual geometrodynamic formula for the fluctuation of the metric\cite{r16},
we have
\begin{equation}
l^2 \delta g_{\imath k} \approx \Delta g_{\imath k} \approx \frac{l_P}{l}\label{e7}
\end{equation}
where $l_P$ is the Planck length.\\
Whence we get
\begin{equation}
\delta g_{\imath k} \sim 1\label{e8}
\end{equation}
if $l$ is of the order $10^{-11}cm$ or the electron Compton wavelength.\\
Similarly using (\ref{e4}) in (\ref{e6}), we recover (\ref{e8}), as in the Weyl
geometry.\\
This establishes the equivalence of the two approaches and reconfirms
the Machian feature, from a more general viewpoint.
\section{The Fine Structure Constant}
In the Weyl geometry considered above, or in a more general scheme of a non
commutative geometry\cite{r17,r18} we have equations like (\ref{e6}), (\ref{e7})
and (\ref{e8}). What they show is that at the Compton wavelength $l$, the
variation of length $\delta l$ is the Planck length, the absolute minimum
length; in this sense $l$ is the minimum physical length (Cf.refs.\cite{r7,r8}and
\cite{r10}).\\
We now observe that, in the above spirit if we consider the gravitational potential
energy of the particle as its inertial energy, we get
\begin{equation}
\frac{NGm^2}{r} \approx mc^2\label{e9}
\end{equation}
In (\ref{e9}) if $N = 1$ and $r \sim$ the Planck length, then we get the
Planck mass. If $N \sim 10^{80}$ and $r \sim$ of the radius of the universe,
we get the mass of an elementary particle like the electron. Thus it is the
presence of $N$ particles that leads to electromagnetism. If $N = 1$, then
we would be at the Planck mass and Planck length. Indeed as can be seen from
(\ref{e1}), in this case the Compton wavelength of such a Planck particle would
equal its schwarzschild radius and electromagnetism and gravitation would become
one (Cf.refs.\cite{r7} and \cite{r19}): electromagnetism would disappear.\\
Let us now examine the above situation in the light of the fluctuational cosmology
model (Cf.ref.\cite{r7,r18} and \cite{r12}). In this model $\sqrt{N}$ particles
are fluctuationally created within the minimum Compton scale time, and as noted
in Section 2 this leads to a consistent explanation of several otherwise adhoc
features as also the recently observed acceleration of the universe. It then
follows that in the early universe when $N \sim 1,$ the present day  Compton wavelenth
would have been of the order of the Planck length.\\
On the other hand the Compton wavelength leads to a correction in the electrostatic
potential experienced by an orbiting electron in an atom, similar to the Darwin
term\cite{r20}. Briefly we have
$$\langle \delta V \rangle = \langle V (\vec r + \delta \vec r)\rangle - V
\langle (\vec r )\rangle$$
$$= \langle \delta r \frac{\partial V}{\partial r} + \frac{1}{2} \sum_{\imath j}
\delta r_\imath \delta r_j \frac{\partial^2 V}{\partial r_\imath \partial r_j} \rangle$$
\begin{equation}
\approx 0(1) \delta r^2 \nabla^2 V\label{e10}
\end{equation}
From (\ref{e10}) it follows that if $\delta r \sim l$, the Compton wavelength
then
\begin{equation}
\frac{\Delta \alpha}{\alpha} \sim 10^{-5}\label{e11}
\end{equation}
where $\alpha$ is the fine structure constant and $\Delta \alpha$ is the change
in the fine structure constant from the early uniiverse.\\
Equation (\ref{e11}) is consistent with the recent observational estimates of
Webb et al., for the evolution of the fine structure constant\cite{r21}.

\end{document}